\documentclass[pra, twocolumn, floatfix,nofootinbib, superscriptaddress]{revtex4-1}
\usepackage{dcolumn}
\usepackage{bm}
\usepackage{graphicx}
\usepackage{amsmath}
\usepackage{enumitem}
\usepackage{latexsym}
\usepackage{amsfonts}
\usepackage{amssymb}
\usepackage{array}
\usepackage{epsfig}
\usepackage[dvipsnames]{xcolor}
\usepackage{todonotes}

\usepackage{color}
\usepackage[colorlinks=true,bookmarks=false,linkcolor=RoyalBlue,urlcolor=RoyalBlue,citecolor=RoyalBlue,breaklinks]{hyperref}
\usepackage{physics}
\usepackage{bbold}

\begin{document}

\title{Adaptive POVM implementations and measurement error mitigation strategies for near-term quantum devices}
\author{Adam Glos$^*$}
\affiliation{Algorithmiq Ltd, Kanavakatu 3 C, FI-00160 Helsinki, Finland}
\author{Anton Nyk\"anen$^*$}
\affiliation{Algorithmiq Ltd, Kanavakatu 3 C, FI-00160 Helsinki, Finland}
\author{Elsi-Mari Borrelli}
\affiliation{Algorithmiq Ltd, Kanavakatu 3 C, FI-00160 Helsinki, Finland}
\author{Sabrina Maniscalco}
\affiliation{Algorithmiq Ltd, Kanavakatu 3 C, FI-00160 Helsinki, Finland}
\author{Matteo A. C. Rossi}
\affiliation{Algorithmiq Ltd, Kanavakatu 3 C, FI-00160 Helsinki, Finland}
\author{Zoltán Zimborás}
\affiliation{Algorithmiq Ltd, Kanavakatu 3 C, FI-00160 Helsinki, Finland}
\author{Guillermo Garc\'{i}a-P\'{e}rez}
\affiliation{Algorithmiq Ltd, Kanavakatu 3 C, FI-00160 Helsinki, Finland}
\date{\today}

\begin{abstract}
%We present new practical methods for implementing POVMs in near term devices, and together with this we also develop novel noise mitigation methods that are built on particular POVM designs.

We present adaptive measurement techniques tailored for variational quantum algorithms on near-term small and noisy devices. In particular, we generalise earlier {\it learning to measure} strategies in two ways. First, by considering a class of adaptive positive operator valued measures (POVMs) that can be simulated with simple projective measurements without ancillary qubits, we decrease the amount of required qubits and two-qubit gates. Second, by introducing a method based on Quantum Detector Tomography to mitigate the effect of noise, we are able to optimise the POVMs as well as to infer expectation values reliably in the currently available noisy quantum devices. Our numerical simulations clearly indicate that the presented strategies can significantly reduce the number of needed shots to achieve chemical accuracy in variational quantum eigensolvers, thus helping to solve one of the bottlenecks of near-term quantum computing.

\end{abstract}
\maketitle

%We first introduce the idea of incorporating Quantum Detector Tomography for noise integration purposes.
%Next, we explain how to use a similar approach while enabling the measurement parameterisation and optimisation.

\section{Introduction}

\def\thefootnote{*}
\footnotetext{These authors contributed equally to this work.}
\def\thefootnote{\arabic{footnote}}

Quantum computers are expected to exceed the capabilities of classical computers for various tasks \cite{montanaro2016quantum}, including optimisation and search \cite{grover1996fast, aaronson2003quantum,brandao2022faster}, linear algebra problems \cite{harrow2009quantum,gilyen2019quantum}, machine learning \cite{biamonte2017quantum,dunjko2020non}, cryptoanalysis \cite{shor1999polynomial,proos2003shor}, and simulations of quantum systems \cite{lloyd1996universal,schuld2019quantum}. However, several of the well-known quantum algorithms require deep circuits and numerous qubits to surpass their classical counterparts. Further, their practical implementation would require quantum error correction, i.e., encoding of the needed logical qubits into a large number of physical qubits in a quantum device where the noise level is below a certain stringent threshold \cite{lidar2013quantum, terhal2015quantum}. Despite the impressive progress made in building proof-of-principle experiments of error correction codes \cite{corcoles2015demonstration, ofek2016extending,krinner2022realizing, acharya2022suppressing} and fault-tolerant operations \cite{egan2021fault}, the road to large-scale fault-tolerant quantum computers is still long. 

%can only be 
%Quantum computing is a rapidly growing multidisci- plinary field with a very clear objective: to understand if, and to what extent, it is possible to build computing machines able to perform tasks that are impossible for con- ventional (classical) computers.
%in many applications. A number of famous algorithms 
%including optimization problems, simulation of quantum many-body systems, and machine learning.

In the near term, the alternative route to useful quantum computation is to develop algorithms adapted for noisy devices and to employ quantum error mitigation methods.
%and run algorithms that are feasible also for noisy devices. 
The most studied group of such algorithms, variational quantum algorithms \cite{cerezo2021variational}, have been developed for optimisation \cite{farhi2014quantum},  machine learning \cite{schuld2019quantum} as well as quantum simulation \cite{peruzzo2014variational}, for which the algorithm is better known as Variational Quantum Eigensolver (VQE).

In implementing simulations of quantum systems, a so-called measurement problem arises from the excessively high cost in terms of the number of measurement shots needed to reconstruct the expectation value of the Hamiltonian. 
In fact, as the size of the problem approaches the regime in which useful quantum advantage over classical methods could be achieved, the traditional measurement approaches lead to unreasonable requirements to reach acceptable accuracy. Many approaches for reducing the amount of required measurement shots have been proposed.For example, methods based on grouping of commuting terms, effective measurement scheduling or optimised qubit tomography have been proposed \cite{Jena2019,Yen2020, huggins2021efficient, gokhale2019minimizing, crawford2021efficient, zhao2020measurement, paini2019approximate,bonet2020nearly,cotler2020quantum, hamamura2020efficient}. To reduce the shots some Pauli strings can be measured simultaneously from the same data set \cite{kandala2017hardware}, or classical machine learning may be used to perform an approximate reconstruction of the quantum state \cite{torlai2018neural, torlai2020precise} or classical shadows of a quantum state \cite{huang2020predicting, hadfield2022measurements, huang2021efficient}. An adaptive method for state tomography using generalised measurements and neural networks was also proposed in \cite{quek2021adaptive}.

A promising adaptive strategy to overcome the measurement problem was presented in \cite{l2m}. There one optimises over a parametric family of informationally complete positive operator-valued measures (IC POVMs) using a hybrid Monte Carlo approach bypassing the need to use tomographic reconstructions of quantum states. The method minimises the statistical errors in the estimation of the target expectation values.
The method was shown to be competitive with state-of-the-art measurement reduction approaches in terms of efficiency. In addition, the informational completeness of the approach offers an advantage, as the measurement data can be reused to infer other quantities of interest that may be used for e.g. post-processing.

The method, however, has two undesirable limitations.
First, the implementation of generic POVMs typically requires at least one ancillary qubit for each qubit to be measured, thus doubling the required size of the quantum hardware.
Second, the the scheme may be sensitive to read-out and gate noise, creating potentially large biases in the energy estimations and hindering the effectiveness of the strategy.
Especially in current quantum hardware, where the qubit number and connectivity are typically limited, and significant noise in the read-out process is expected, the above mentioned problems become paramount.
In this paper we address these limitations and develop schemes for both implementing dilation free POVMs and read-out noise mitigation strategies suitable for adaptive measurement schemes. The two schemes can be implemented independently.
We demonstrate that the proposed methods
significantly reduces the number of needed shots to achieve chemical accuracy in current hardware.are effective on current hardware in significantly reducing the number of needed shots to achieve chemical accuracy in VQE. The results therefore indicate that adaptive measurement strategies, combined with qubit-efficient measurements and appropriate noise mitigation, can help remove one of the critical bottlenecks of near-term quantum computing.

The paper is structured as follows. In section \ref{sec:dilation_free}, we first discuss how to implement dilation-free POVMs and how they may be parameterised for adaptive measurement strategies.
In section \ref{sec:QDT}, we present a strategy for read-out noise mitigation using quantum detector tomography, that may be used for adaptive POVM implementations with and without dilation. We further demonstrate the effectiveness of the methods on simulated hardware and finally, in section \ref{sec:conclusions}, present the conclusion and outlook.

\section{Implementation of dilation-free POVMs}
\label{sec:dilation_free}

In this section, we discuss the class of  dilation free POVMs, which in the literature are also referred to as Projective Measurement (PM)-Simulable measurements \cite{Haapasalo2011,oszmaniec2017}.
%The class of POVMs that we make use of is referred to in the quantum information literature as Projective Measurement (PM)-Simulable measurements \cite{Haapasalo2011,oszmaniec2017}.
Essentially, these measurements combine projective measurements with classical randomness (probabilistic mixing) and classical post-processing.
In other words, by combining projective measurements and classical resources, one can implement this large class of non-projective measurements, including a subset of IC-POVMs, without the need for ancillary qubits.
\subsection{Physical implementation for a single qubit}
\label{sec:method_description}

\begin{figure}[h]
    \includegraphics[width=.95\columnwidth]{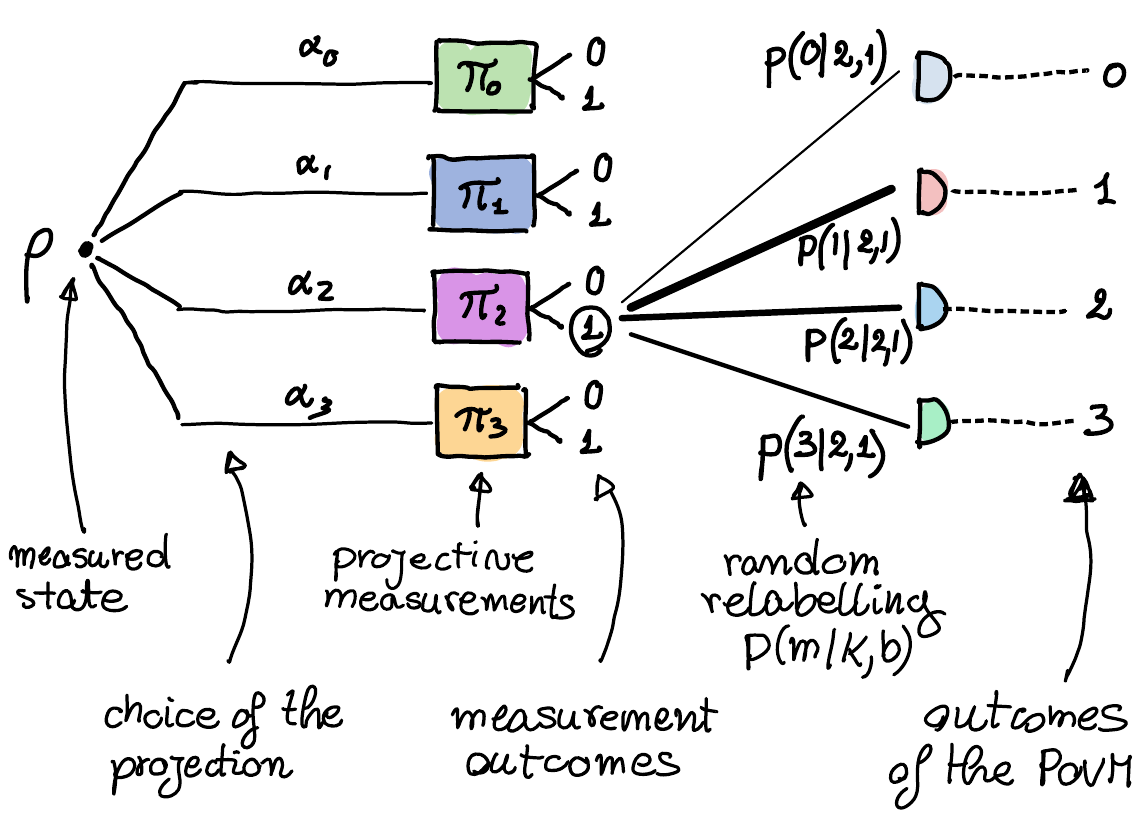}
    \caption{Schematic description of a single-qubit PM-Simulable POVM. For each preparation of the state $\rho$, a projective measurement $\pi_k = \lbrace \pi_{k,0}, \pi_{k,1} \rbrace$ is chosen randomly with probability $\alpha_k$. After the measurement has been performed, the outcome $(k,b)$ is post-processed by relabelling it to one of the POVM outcomes $m$ with probability $P(m|k, b)$. In the figure, the post-processing of outcome $b=1$ of the projective measurement $\pi_2$ is highlighted with solid lines, the width of which indicates the value of the corresponding probability. Hence, upon obtaining outcome $(k=2, b=1)$ in the first stages of the procedure, a number $m \in \{0, 1, 2, 3\}$ is sampled from the probability distribution $P(m | 2, 1)$ and the outcome is relabelled accordingly.}
    \label{fig:pm-simulated_povm}
\end{figure}

The basic structure of PM-Simulable POVMs for the case of a single qubit is given as follows: Consider a collection of $K$ projective measurements $\lbrace \pi_{k,0}, \pi_{k,1} \rbrace_{k =0}^{K-1}$ (i.e, for each of the POVMs, labelled by $k=0, \ldots K{-}1$, one has that  $\pi_{k,b}^2 = \pi_{k,b}$, for $b=0,1$,  and $\pi_{k,0} + \pi_{k,1} = \mathbb{I}$) and a $K$-outcome probability distribution $\lbrace \alpha_k \rbrace_{k =0}^{K-1} $ (i.e., with $\alpha_k \geq 0$ and $\sum_k \alpha_k = 1$).
One can now sample using this probability distribution a value of $k = 0, \dots, K{-}1$, and then apply the $k$-th measurement $\lbrace \pi_{k,0}, \pi_{k,1} \rbrace$. The overall process is a measurement with $2 K$ outcomes described by the effects $\lbrace \alpha_k \pi_{k,b} \rbrace_{k =0 , b =0}^{K-1, 1}$, in the sense that the probability of obtaining outcome $(k, b)$ is $p_{k,b} = \mathrm{Tr} [ \alpha_k \pi_{k,b} \rho ]$, where $\rho$ is the state of qubit.
In addition, one can post-process the outcomes randomly by relabelling them according to some conditional probability distribution $P (m | k, b)$.
In other words, after obtaining some outcome $(k, b)$ from the former $2K$-outcome measurement, we classically draw a value of $m$ from a probability distribution $P (m | k, b)$, different for every outcome $(k, b)$, and use the resulting value of $m$ to relabel the outcome.
The probability of obtaining outcome $m$ in the process is thus given by $p_m = \sum_{k, b} P(m | k, b) p_{k,b} = \sum_{k, b} P(m | k, b) \mathrm{Tr} [ \alpha_k \pi_{m,b} \rho ]$, so the $m$-th outcome has an associated effect $\Pi_m = \sum_{k, b} P(m | k, b) \alpha_k \pi_{k,b}$.
Hence, by combining a set of PMs with classically random processes, one can construct non-projective POVMs.
The process is summarised in Fig.~\ref{fig:pm-simulated_povm}.

Let us note that quantum computers are typically designed to enable the measurement of their qubits in the computational basis, that is, ideally in terms of the projectors $| 0 \rangle \langle 0|$, and $| 1 \rangle \langle 1|$.
Moreover, single-qubit gates are typically easy to implement with high fidelity.
These two features combined allow us to construct an arbitrary single-qubit projective measurement $\pi_{k,b}$ by applying an appropriate single-qubit gate $U_k$ to the qubit before the computational basis measurement. Indeed, by choosing a $U_k$ that satisfies  $\pi_{k,b} = U_k^\dagger | b \rangle \langle b| U_k$ for $b=0,1$, outcome $b$ is obtained with probability $p_b = \mathrm{Tr} [U_k \rho U_k^\dagger | b \rangle \langle b|] =  \mathrm{Tr}[\rho U_k^\dagger | b \rangle \langle b| U_k] =\mathrm{Tr} [\rho \pi_{k,b}]$.

In summary, given a set of $K$ single-qubit unitaries $\lbrace U_k \rbrace_{k =0 }^{K-1}$, a probability distribution $\lbrace \alpha_k \rbrace_{k =0 }^{K-1}$, and a set of $M$-outcome conditional probability distributions $\lbrace \lbrace P(m | k, b)\rbrace_{m=0}^{M-1} \rbrace_{k =0 , b =0}^{ K-1, 1}$, we can construct an $M$-outcome POVM through the following procedure:
\begin{itemize}
    \item[1.] Sample a value of $k$ with probability $\alpha_k$.
    \item[2.] Apply the single-qubit unitary $U_k$ on the qubit.
    \item[3.] Measure the qubit in the computational basis to obtain an outcome $b \in \lbrace 0, 1 \rbrace$.
    \item[4.] Sample a value of $m$ with probability $P(m | k, b)$, and relabel the outcome using the resulting $m$.
\end{itemize}
The POVM is described by the effects
\begin{equation}\label{eq:effects}
    \left\lbrace \Pi_m = \sum_{k = 0}^{K-1} \sum_{b = 0}^{1} P(m | k, b) \alpha_k U_k^\dagger | b \rangle \langle b| U_k \right\rbrace_{m =0}^{M-1},
\end{equation}
i.e., the probability for the process to yield outcome $m$ when applied to a qubit in state $\rho$ is $\mathrm{Tr} [\rho \Pi_m]$, with $\Pi_m$ given above.
Finally, notice that $\sum_m \Pi_m = \mathbb{I}$ and $\Pi_m$ is positive semidefinite for any $m$, so the set in Eq.~\eqref{eq:effects} is a valid POVM.

\subsection{Parameterisation of ancilla-free POVMs and their use in an adaptive measurement scheme}
\label{sec:parameterisation}

We now turn to the parameterisation of dilation free POVMs described in Section \ref{sec:method_description} through the characterisation of  
the corresponding POVM effects.
If the corresponding POVM is IC, the outcomes can be used to reconstruct expectation values of observables using Eq.~(2) in Ref.~\cite{l2m}. More precisely, since the single-qubit effects can be calculated explicitly on a classical computer using Eq.~\eqref{eq:effects}, we can use them to calculate the coefficients $b_{im}$ fulfilling $\sigma_i = \sum_m \beta_{im} \Pi_m$, where $\sigma_i$ are Pauli matrices, as in Ref.~\cite{l2m}, and the hybrid quantum-classical Monte Carlo method can be used with this POVM class as well.
Therefore, it is important to ensure that each single-qubit POVM used is IC, meaning that it contains a subset of four linearly independent effects.

Notice that each single-qubit POVM depends on several elements that can be modified:
the set of $K$ single-qubit unitaries $\lbrace U_k \rbrace_{k=0}^{K-1}$, the probability distribution $\lbrace \alpha_k \rbrace_{k=0}^{K-1}$, and the set of $2K$ $M$-outcome conditional probability distributions $\lbrace \lbrace P(m | k, b)\rbrace_{m=0}^{M-1} \rbrace_{k =0 , b =0}^{ K-1, 1}$.
Each of the elements may be parametrised by choosing them from a specific class or family.

For instance, the probability distribution $\lbrace \alpha_k \rbrace_{k=0}^{K-1}$ is given by a collection of $K$ positive numbers adding up to one, so they can be associated with the squares of the coordinates of points on the surface of a $(K-1)$-dimensional sphere as follows: let $\{ \varphi_i \in [0, \pi / 2] \}_{i=0}^{K-2}$ be a set of $K-1$ angles.
They uniquely identify a point in a hyperquadrant of a unit-radius $(K-1)$-dimensional hypersphere in $\mathbb{R}^k$ with euclidean coordinates given by
%such that its Euclidean coordinates in $\mathbb{R}^K$, given by
\begin{equation}\label{eq:hypercoords}
\begin{aligned}
x_0 &= \cos(\varphi_0)\\
x_1 &= \sin(\varphi_0)\cos(\varphi_1)\\
& \; \; \vdots \\
x_{K-2} &= \sin(\varphi_0)\sin(\varphi_1)\cdots\cos(\varphi_{K-2})\\
x_{K-1} &= \sin(\varphi_0)\sin(\varphi_1)\cdots\sin(\varphi_{K-2})
\end{aligned},
\end{equation}
which are all positive numbers, given that $\cos(\varphi_i) > 0$ and $\sin(\varphi_i) > 0$ for all $i$.
The set of numbers $\{ \alpha_i = x_i^2 \}_{i=0}^{K-1}$ is composed of positive numbers adding up to identity, so it defines a $K$-outcome probability distribution. Moreover, given any $K$-outcome probability distribution, Eq.~\eqref{eq:hypercoords} can be inverted to find the corresponding set of angles.
This procedure thus defines a bijection between the space of $K$-outcome probability distributions and positive-coordinate quadrants of $(K-1)$-dimensional spheres, so we can use sets of angles to parameterise distributions.

The same technique can be used to parameterise each of the probability distributions in the relabelling.
In that case, for each possible outcome $(k, b)$, the final measurement outcome $m$ is decided randomly with probability $P(m | k, b)$, so we need one such $M$-outcome probability distribution for each outcome $(k, b)$ (that is, $2 K$ $M$-outcome probability distributions in total).
In this case, this means that we need to define a set of $M-1$ angles for every outcome $(k, b)$.
Each such set uniquely identifies a set of $M$ positive numbers $\{ x^{(k,b)}_i \}_{i=0}^{M-1}$, so we can set $\lbrace P(m | k, b) = (x^{(k,b)}_{m})^2\rbrace_{m=0}^{M-1}$. It is important to stress that other parameterisations of discrete probability distributions are possible.

Single-qubit unitaries, on the other hand, are determined by three parameters $\phi_0$, $\phi_1$, and $\phi_2$,
\begin{equation}
    U (\phi_0, \phi_1, \phi_2) = 
    \begin{pmatrix}
    \cos (\phi_0 / 2) & - e^{i \phi_1} \sin (\phi_0 / 2) \\
    e^{i \phi_2} \sin (\phi_0 / 2) & e^{i (\phi_1 + \phi_2)} \cos (\phi_0 / 2)
    \end{pmatrix}.
\end{equation}
Consequently, each single-qubit POVM depends on the parameters corresponding to the probability distributions as well as the unitaries. 
Moreover, since this applies to each single-qubit POVM and the effects of the joint $N$-qubit POVM are simply products of them, the joint effects depend on the set of all single-qubit parameters.
Collecting all these parameters in a vector $\vec{\varphi}$, we can write the resulting $N$-qubit POVM in terms of the effects $\lbrace \Pi_\mathbf{m} (\vec{\varphi}) \rbrace$, where $\mathbf{m}$ represents an length-$N$ outcome string, as in Ref.~\cite{l2m}.
The adaptive algorithm from Ref.~\cite{l2m} is agnostic to the specific class of POVM under use, as long as it can be constructed in terms of parameterised single-qubit IC-POVMs; thus, the algorithm can be directly applied with PM-Simulable measurements.

The methodology introduced above contains the basic elements to physically implement the adaptive measurement strategy with PM-Simulable measurements. As near-term quantum devices are mostly available through the cloud, in Appendix \ref{sec:reduce_latency} we describe a method to reduce the total execution time when efficient communications protocols between the classical and the quantum computers (such as Qiskit runtime) are not present.

%Figure~\ref{fig:results} shows the performance of the adaptive measurement scheme for VQE for H$_2$ molecule mapped into a 4-qubit Hamiltonian using the Parity fermion-to-qubit mapping \cite{bravyi2017tapering}. Both the estimated error and the actual error in the estimation decrease significantly faster than $S^{-1/2}$, where $S$ is the total number of shots, showing that the proposed POVM implementation can be optimized to improve the measurement precision.

%\begin{figure}[h!]
%    \centering
%    \includegraphics[width=0.9\columnwidth]{H2_adaptive.pdf}
%    \caption{Estimation of the energy of the H$_2$ molecule mapped into a 4-qubit Hamiltonian using the Parity fermion-to-qubit mapping. The state is the approximated ground state obtained after a VQE run. The solid line shows the error of the estimation as a function of the total number of measurement rounds $S$ averaged over 100 realisations. The dashed line shows the estimated error. The dotted line indicates the scaling $S^{-1/2}$ corresponding to a non-adaptive measurement scheme. Using the method here introduced, the estimation approaches the exact value significantly faster than $S^{-1/2}$.}
%    \label{fig:results}
%\end{figure}

\begin{figure*}[t]
\includegraphics[width=0.8\textwidth]{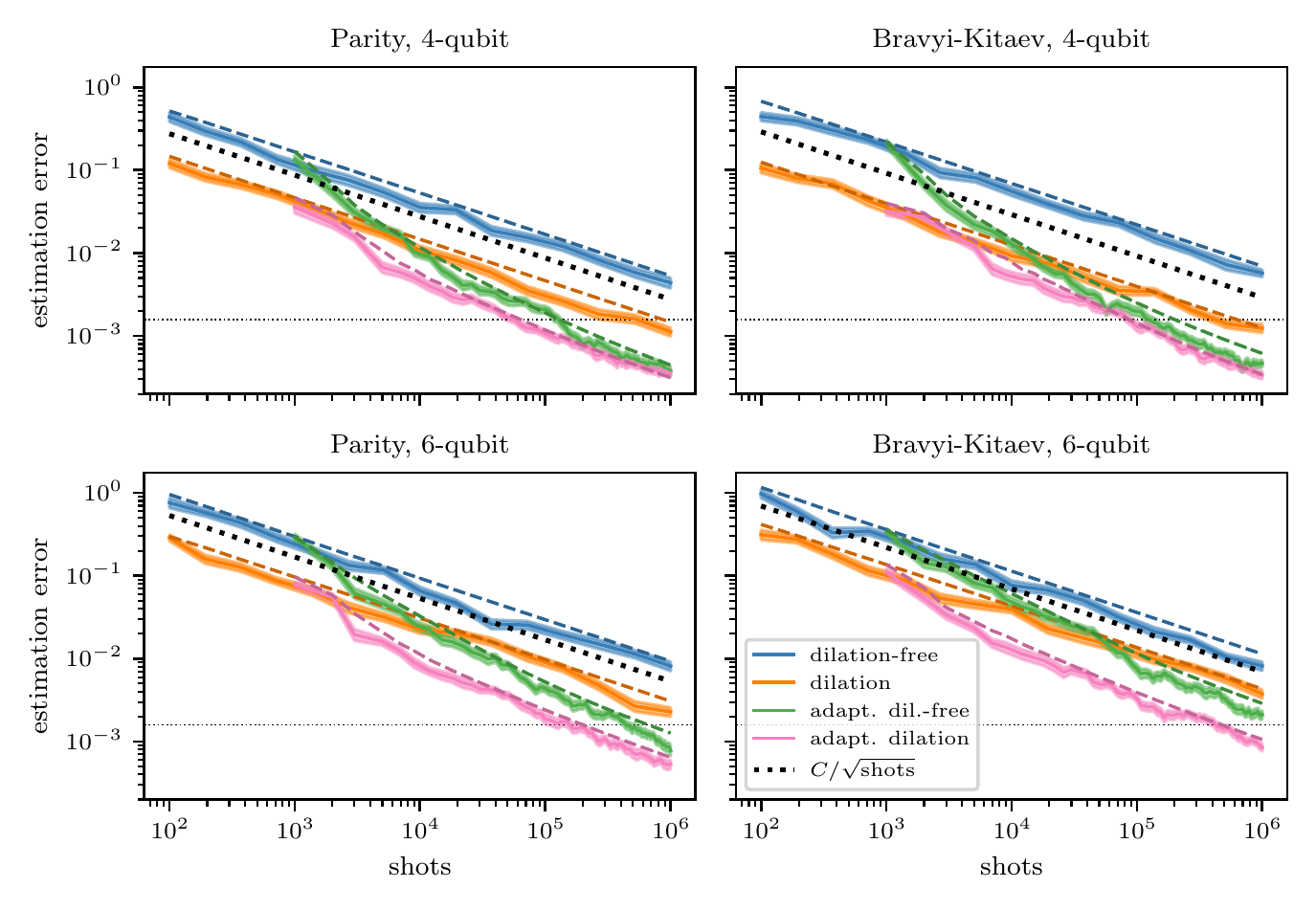}
\caption{VQE ground state energy estimation error for a simulation of $H_2$ molecule with bond length $0.75\textup{~\AA}$. The computations where performed on a simulated fully connected device without noise. We present results for parity preserving (left) and Bravyi–Kitaev (right) fermion-to-qubit mappings \cite{mappings} for four (up) and six (down) qubits using STO-3G and 6-31G basis, respectively. The orange (blue) curve shows the true mean error among 40 estimations for a measurement performed with (without) dilation. The green (pink) curve shows the corresponding true mean error for the adaptive measurement scheme. The corresponding areas represent the standard error w.r.t. true energy of the ground state among 40 runs taken. The dashed lines give the estimated standard errors, that is, computed without knowledge of the true ground energy.  The horizontal dashed line corresponds to chemical precision $1.59 mH$, while the oblique dotted line is a reference line $\frac{C}{\sqrt{S}}$ representing a typical estimation error precision decrease with respect to the given number of shots $S$.}
\label{fig:no_noise}
\end{figure*}

In Fig.~\ref{fig:no_noise} we plot the $\rm{H_2}$ molecular ground state energy estimation error for a state obtained with Qubit-ADAPT-VQE \cite{q_adapt} for different fermion-to-qubit mappings and different basis sets. The simulations were performed in the absence of noise. We compare the adaptive measurement scheme to the non-adaptive case for both a generalised measurement performed via dilation and a PM-simulable POVM. We see that in all the cases considered here the adaptive strategies clearly outperform the standard approach. Further, when we limit to the PM-simulable POVMs in the adaptive scheme, the number of measurement shots required to reach chemical precision increases, but remains nevertheless substantially smaller than for the non-adaptive strategies.

\section{Measurement noise mitigation via quantum detector tomography}
\label{sec:QDT}

Quantum computers and quantum simulators are subjected to read-out errors. The physical measurement process is generally faulty and the realised measurement operator may differ significantly from the idealised one. For example, in NISQ devices, the measurement in the computational basis of state $\ket{0}$ ($\ket{1}$) can yield outcome $1$ ($0$) with non-zero probability. To obtain reliable results from any computation performed on the device, such measurement errors must be addressed.
Generally, read-out noise mitigation strategies aim at correcting the empirical distribution of outcomes by modelling the measurement error. In these approaches one considers often only projective measurements and stochastic errors (or bit flips) as specific error models \cite{Maciejewski2020mitigationofreadout,Chen2019,geller2020rigorous,maciejewski2021modeling,bravyi2021mitigating,nation2021scalable,dahlhauser2021modeling,funcke2022measurement}, although sometimes also full detector tomography is performed \cite{Maciejewski2020mitigationofreadout}.

Here, we propose to utilise informationally complete POVMs (IC-POVMs) to devise an effective strategy for read-out noise mitigation.
Such generalised measurements require, in addition to the final ideally projective measurement, some intermediate physical operations, such as e.g.\ implementing additional gates with ancillary qubits.
When implemented on a real device, these operations are also imperfect, which can introduce additional measurement errors.
As we now show, it is nevertheless possible to use hybrid quantum-classical Monte Carlo methods for observable  averaging.

Let us first decompose each single-qubit operator appearing in the representation of any many-body observable $\mathcal{O}$ in terms of the POVM effects.
For example, given an operator as a linear combination of Pauli strings, $\mathcal{O} = \sum_\mathbf{k} c_\mathbf{k} P_\mathbf{k}$, where  $P_\mathbf{k} = \bigotimes_{i=1}^{N} \sigma_{k_i}^{(i)}$ and  $\sigma_0^{(i)} = \mathbb{I}^{(i)}, \sigma_1^{(i)} = \sigma_x^{(i)}, \dots$ are Pauli operators acting on each qubit $i$, we can express these in terms of the effects $\Pi_m^{(i)}$ as $\sigma_{k}^{(i)} = \sum_m b_{km}^{(i)} \Pi_{m}^{(i)}$, with which we can write
\begin{equation}\label{eq:decomposition}
\begin{aligned}
\mathcal{O} &= \sum_\mathbf{k} c_\mathbf{k} \bigotimes\limits_{i=1}^{N} \sigma_{k_i}^{(i)} 
= \sum_\mathbf{k} c_\mathbf{k} \bigotimes\limits_{i=1}^{N} \left(\sum\limits_{m_i = 0}^{3} b_{k_i m_i}^{(i)} \Pi_{m_i}^{(i)} \right) \\
&= \sum_\mathbf{m} \sum_\mathbf{k} c_\mathbf{k} \prod\limits_{i=1}^{N} b_{k_i m_i}^{(i)} \Pi_\mathbf{m} \equiv \sum_\mathbf{m} \omega_\mathbf{m} \Pi_\mathbf{m}.
\end{aligned}
\end{equation}
The expectation value of the operator is then
\begin{equation}\label{eq:povm_exp_val}
\langle \mathcal{O} \rangle = \mathrm{Tr} [ \rho \mathcal{O} ] = \sum_\mathbf{m} \omega_\mathbf{m} \mathrm{Tr} [ \rho \Pi_\mathbf{m} ] = \sum_\mathbf{m} \omega_\mathbf{m} p_\mathbf{m},
\end{equation}
where $p_\mathbf{m}$ is the probability to obtain outcome $\mathbf{m}$.
An obvious consequence of noise in the device translates into the possibility that the effects describing the actual physical measurement are different from the idealised ones expected in the absence of noise.
Thus, if one relies on the idealised effects in the above computation, that is, one calculates the $b_{k_i m_i}^{(i)}$ such that $\sigma_{k}^{(i)} = \sum_m b_{km}^{(i)} \Pi_{\mathrm{ideal}, m}^{(i)}$, the result will generally be biased, since the actual probability distribution of the outcomes is affected by the noise and hence can be different from the idealised one $\{ p_{\mathrm{ideal}, \mathbf{m}} \}$.

Using the effects that describe the actual physical measurements $\{ \Pi_{\mathrm{real}, m}^{(i)} \}$, that is, using the coefficients $b_{k_i m_i}^{(i)}$ corresponding to the decomposition of the Pauli operators in terms of these effects in Eq.~\eqref{eq:povm_exp_val}, would remove such bias, given that, by definition, they describe the actual probability distribution of the outcomes $\{ p_{\mathrm{real}, \mathbf{m}} \}$ (assuming the real measurement process can be regarded as independent processes for every qubit, i.e., there are negligible cross-talk effects and spurious correlations).
In a real device, the actual measurement $\{ \Pi_{\mathrm{real}, m}^{(i)} \}$ is however unknown.
We propose using Quantum Detector Tomography (QDT) to approximate them from empirical data.

Multiple methods to perform QDT have been proposed in the literature \cite{Sanchez-Soto1999, Fiurasek2001, Dariano2004, Lundeen2008, Maciejewski2020mitigationofreadout}. These methods require preparing the subsystems, e.g.\ the qubits, in a set of initial states and measuring them with the measurement apparatus that one is interested in characterising.
One can then collect outcome statistics and write a system of equations for the POVM effects describing the physical measurement, that is, to determine the effects that reproduce such outcome statistics given the initial states.
If the set of initial states contains enough states, the system of equations becomes determined and the solution is unique.
In practice, due to finite statistics, one needs to resort to more robust numerical techniques for the reconstruction of the effects, such as maximum likelihood estimation~\cite{Fiurasek2001}.

In general, noisy devices may not enable perfect state preparation, which can hinder QDT.
However, as explained in Ref.\ \cite{Maciejewski2020mitigationofreadout}, Sect.\ 6.1, the initial states required for QDT can be prepared by applying a single-qubit unitary on each qubit.
Single-qubit unitaries are usually high-quality in any potentially useful quantum computer, so one can assume the state preparation quality to be sufficient for QDT.
In addition, certain techniques such as randomised benchmarking enable estimating the quality of these unitaries discounting the effect of the noise in the measurement apparatus, so it is even possible to verify whether QDT will yield a reliable characterisation of the apparatus.
Applications of QDT to near-term quantum devices are usually aimed at performing error mitigation \cite{Maciejewski2020mitigationofreadout}.

%The basic working principle of QDT consists in preparing a collection of states and measuring them to infer the empirical outcome probability distribution on every subsystem (e.g. qubit).
%Once these are known, one proceeds to find the POVM effects $\{ \Pi_{\mathrm{QDT}, m}^{(i)} \}$ that best describe the observed probability distributions.
%This is often done following a likelihood maximisation approach.

Using numerical simulations in which the $\{ \Pi_{\mathrm{real}, m}^{(i)} \}$ can be fully controlled and known, it can be shown that the resulting effects approximate well the actual ones, that is, $ \Pi_{\mathrm{QDT}, m}^{(i)} \approx \Pi_{\mathrm{real}, m}^{(i)}$.
Thus, the method proposed in this section is simply to use QDT to determine $\{ \Pi_{\mathrm{QDT}, m}^{(i)} \}$ and then use these effects in the hybrid quantum-classical Monte Carlo from Ref.\ \cite{l2m}.
In other words, the strategy is to compute the coefficients $\{ b_{km}^{(i)} \}$ fulfilling $\sigma_{k}^{(i)} = \sum_m b_{km}^{(i)} \Pi_{\mathrm{QDT}, m}^{(i)}$ for all Pauli operators $\sigma_{k}^{(i)}$ once the POVM effects have been determined via QDT for each qubit $i$ and use them to compute $\omega_{\textbf{m}} = \sum_\mathbf{k} c_\mathbf{k} \prod_{i=1}^{N} b_{k_i m_i}^{(i)}$ for each $N$-qubit outcome $\textbf{m}$ obtained upon measurement of state $\rho$.
The expectation values of observables, $\langle \mathcal{O} \rangle$, can then be estimated as $\bar{\mathcal{O}} = \sum_{s = 1}^{S} \omega_{\textbf{m}_s} / S$, where $S$ is the total number of shots and $\textbf{m}_s$ is the $s$-th POVM outcome.
Similarly, the statistical error of the estimator $\bar{\mathcal{O}}$ can be estimated as $\epsilon = \sqrt{(\sum_{s = 1}^{S} \omega_{\textbf{m}_s}^2 / S - \bar{\mathcal{O}}^2) / (S - 1)}$.

Notice that this method is implementation-agnostic.
Regardless of the class of POVMs and the additional quantum or classical resources they require, we always infer effects that live in the Hilbert space of each subsystem (e.g.\ qubit).
Also, we note that this QDT-based noise mitigation strategy was successfully employed in Ref.~\cite{fischer2022ancilla} for qudit-based POVM implementations on IBM devices.

\subsection{Measurement noise mitigation for adaptive measurement strategies}
\label{sec:QDT_adaptive}
The approach presented above cannot be used for adaptive measurement strategies relying on parametric families of POVMs presented in Ref.~\cite{l2m}. Indeed, in such methods given the data obtained with some IC POVM, the performance of other POVMs in the parametric family are classically estimated in order to propose a different, more efficient, generalised measurement for the next set of measurements. Since the effects of the POVM are determined  a posteriori via QDT, the adaptive strategy cannot be straightforwardly used.

For such adaptive strategies, a possible solution is to model the noisy detector, that is, to construct a classical model that associates (or predicts) realistic POVM effects to any value of the POVM parameters.
This involves, in general, to characterise each element in the physical implementation of the generalised measurement.

For instance, generalised measurements implemented on quantum devices generally involve a measurement native to the device.
In the case of quantum computers, these may be aimed at measuring the qubits in the computational basis.
However, due to noise and other imperfections, as explained above, the actual measurement is not given by the ideal projectors $\ket{0} \bra{0}$ and $\ket{1} \bra{1}$.
Instead, the real physical measurement is described by a two-outcome POVM with unknown effects $\{ M_{\mathrm{real}, 0}, M_{\mathrm{real}, 1} \}$.
As in the previous section, these effects can be approximated by means of QDT, which yields $\{ M_{\mathrm{QDT}, 0}, M_{\mathrm{QDT}, 1} \}$.
Note that, in this case, QDT is used to determine only one part of the whole POVM implementation: the final measurement.

In addition, generalised measurements typically involve additional operations, which may even rely on the interaction between the logical systems (e.g.\ qubits) and some additional degrees of freedom, be it ancillary qubits or excited states of the physical qubit \cite{IBM_ancilla_free}.
These operations are generally designed to be unitary, and they typically are the elements within the POVM implementation that can be modified and, hence, enable the parameterisation of the POVM. In some cases, these operations are implemented as a sequence of operations that are native to the device, some of which may be parameter-free (e.g.\ CNOT gates). 

Due to noise, they result in non-unitary transformations that can be described in terms of  completely positive and trace preserving (CPTP) maps, i.e. quantum channels.The parameter-free operations can be characterised through Quantum Process Tomography (QPT) in order to reconstruct their Choi matrix or equivalent Kraus decomposition, so that we can classically calculate their effect on arbitrary states $\rho$ as $\mathcal{E}_{\mathrm{QPT}}^{\mathrm{param-free}} (\rho)$.

Other operations are implemented by means of pulses of a certain duration.
These dynamics are idealised as being generated by a Hamiltonian, hence resulting in a unitary transformation.
In practice, the dynamics is better described by a master equation in which the dissipative terms account for the effect of noise.
The corresponding master equation can be inferred experimentally, so that one can model the dynamics and hence classically assess what channel $\mathcal{E}_{U_{\vec{z}}}$ is actually implemented when intending to perform the unitary $U_{\vec{z}}$ parameterised by $\vec{z}$.
In the literature of quantum computing, simpler models for noisy gates are commonly found.
For instance, it is customary to assume that the result of executing a given pulse is the ideal unitary followed by some quantum channel, e.g., depolarising.
Such other noisy gate models can be employed here too.
In fact, the methodology explained here only assumes the use of some noisy gate model, so that the realistic effect of executing parameterised gates can be accounted for, but remains agnostic to the particular model.

These operations may involve other subsystems, the state of which must be known.
For instance, the generalised measurement might be based on the interaction with ancillary qubits in some state, such as $\ket{0}$.
In real situations, the state of the ancilla may have been perturbed and hence be different from the expected one.
To determine it, one may use single-qubit Quantum State Tomography (QST) after running all other operations necessary for the computation (e.g.\ the ansatz in VQE), since these may be responsible for the perturbation.
Doing so requires measuring the ancillary qubits in a tomographically complete basis.
In this way, the state of a given ancilla can be assumed to be $\rho_{\mathrm{QST}}$ instead of the idealised state, and this can be incorporated in the model of the detector as well.

Once all these pieces have been determined, the family of noisy, not idealised, POVM can be used in the following way.
Let us suppose that we are using some parametric family of POVMs, and we consider a specific instance corresponding to parameters $\vec{x}$.
Then, according to our classical model of the detector, based on the above considerations, we can assume that the probability $p_i$ for outcome $i$ to occur on state $\rho$ is $p_i = \mathrm{Tr}[\mathcal{E}_{\vec{x}} (\rho \otimes \rho_{\mathrm{QST}}) M_{\mathrm{QDT}, i}]$, where
$M_{\mathrm{QDT}, i}$ is the effect describing the outcome $i$ of the ideally projective measurement of the device, as determined by means of QDT, $\mathcal{E}_{\vec{x}}$ is the quantum channel resulting from the composition of all the intermediate channels, both parametric and parameter-free, which can be calculated classically.
In addition, it also accounts for the initial noisy state of the ancilla as determined by QST in the cases in which ancillary subsystems are involved.
% In general, however, we shall consider this channel as the classical description or model of the physical transformation of the system in the Hilbert space involved throughout the generalised measurement process.

\begin{figure*}[t]
    \includegraphics[width=0.8\textwidth]{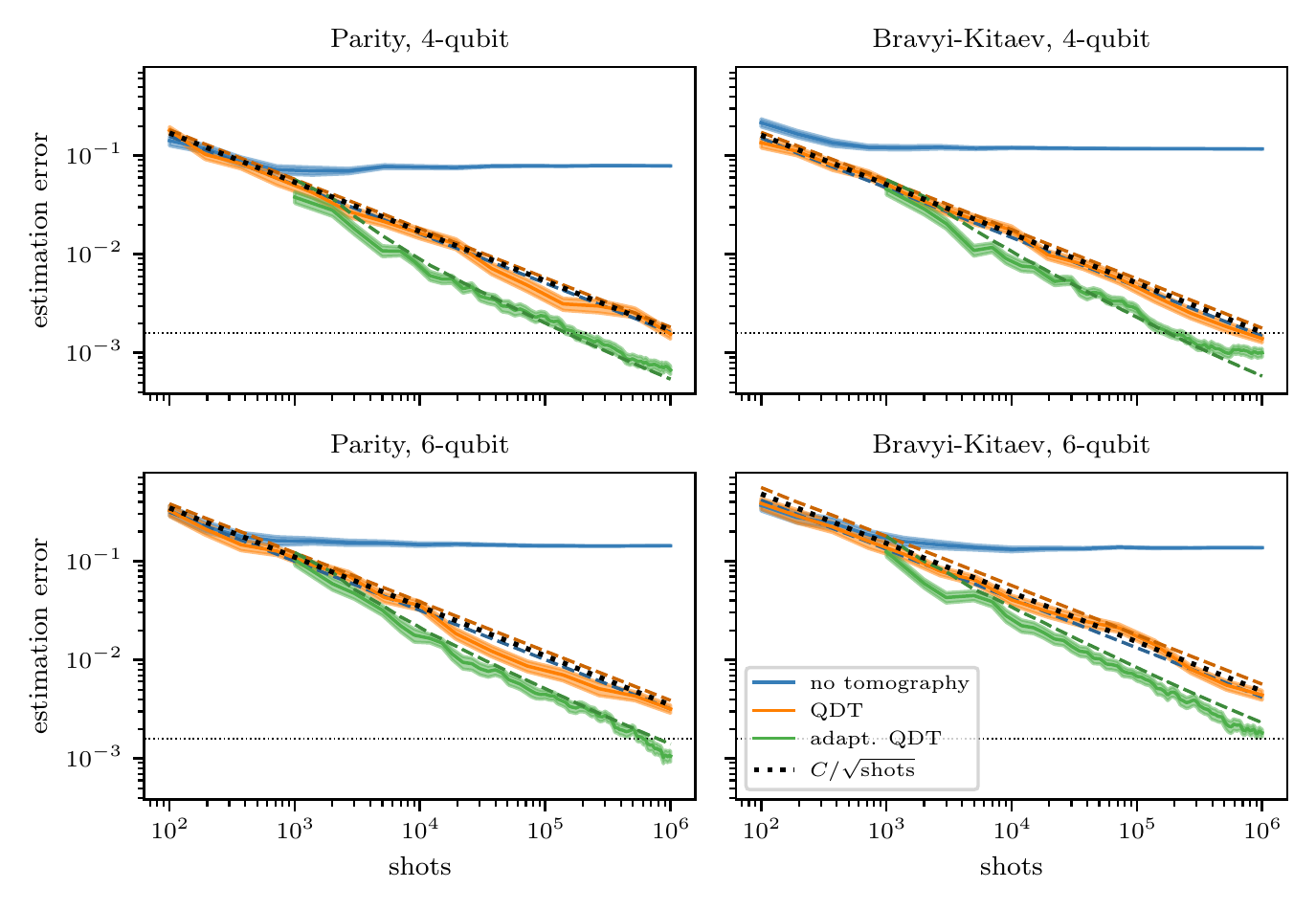}
    \caption{VQE ground state energy estimation error for a simulation of $H_2$ molecule with bond length $0.75\textup{~\AA}$, when measurements are performed with a dilation doubling the amount of qubits required for the computation. The computations where performed on a simulated fully connected device, for which the noise parameters were randomly picked from IBM Hanoi system. We present results for parity preserving (left) and Bravyi–Kitaev (right) fermion-to-qubit mappings \cite{mappings}, four (up) and six (down) qubits in STO-3G and 6-31G basis, respectively. The blue solid curve gives the true mean error among 40 estimations when no readout noise mitigation is employed. The orange and green solid curves give the true mean errors when QDT is used to correct the bias in the non-adaptive and adaptive strategies respectively. The corresponding areas represent the standard error w.r.t. true energy of the ground state among 40 runs taken. The dashed line gives the standard error w.r.t. estimated energy. The horizontal dotted line represent the chemical accuracy $1.59 mH$, while the oblique dotted line is a reference line $\frac{C}{\sqrt{S}}$ representing a typical estimation error precision decrease with respect to the given number of shots $S$. }
    \label{fig:fake_hanoi_dilation}
\end{figure*}

With all these ingredients at hand, we can classically calculate the effects, $\{ \Pi_{\mathrm{model}, i} (\vec{x}) \}$, fulfilling
\begin{equation}\label{eq:effect_model}
    \mathrm{Tr}[ \rho \Pi_{\mathrm{model}, i} (\vec{x})] = \mathrm{Tr}[\mathcal{E}_{\vec{x}} (\rho \otimes \rho_{\mathrm{QST}}) M_{\mathrm{QDT}, i}].
\end{equation}
The general solution to this equation is presented in Appendix \ref{sec:ModelEffects}. The reconstructed effects should describe the real outcome probability distributions reasonably well.
In this way, we can construct parametric families of POVMs that naturally incorporate the noise and imperfections of the device.

Since the effects can be calculated on a classical computer before implementing them physically, the adaptive strategy from Ref.\ \cite{l2m} can be used.
In practice, it may be beneficial to use the modelled effects $\{ \Pi_{\mathrm{model}, i} (\vec{x}) \}$ only to find better parameters for the next iteration, and once these parameters have been chosen, to obtain more accurate effects by means of QDT as in the previous section.

For the single-qubit dilation POVM, used in Ref.\ \cite{l2m}, several sources of noise must be considered.
On the one hand, at the time of the measurement, the state of the ancilla may have been driven away from the ground state, and be something else.
As explained above, we can use QST to determine it approximately.
A second source of noise is the lack of unitarity throughout the system-ancilla interaction.
In the IBM devices, for example, the unitary $U_{\vec{z}}$ is decomposed in terms of a sequence of single-qubit unitaries and CNOTs.
Characterising the CNOTs using QPT, and the single-qubit dynamics by means of an appropriate master equation or some simple model, we can approximate $\mathcal{E}_{\vec{x}}$ as $\mathcal{E}_{\vec{x}} = \mathcal{E}_{\vec{z}_1} \circ \mathcal{E}_{\vec{z}_2} \circ \mathcal{E}_{\mathrm{QPT, CNOT}} \circ \cdots$, where the sequence of compositions follows that of the physical implementation.
Finally, the measurement may be imperfect as well, so QDT can be employed to determine the four-outcome POVM $\{ M_{\mathrm{QDT}, 0}, \ldots, M_{\mathrm{QDT}, 3} \}$ to substitute the idealised two-qubit computational basis projectors.
With this information, we may assume that the probability for outcome $i$ to occur given state $\rho$ is $p_i = \mathrm{Tr}[\mathcal{E}_{\vec{x}} (\rho \otimes \rho_{\mathrm{QST, ancilla}}) M_{\mathrm{QDT}, i}]$, instead of $p_i = \mathrm{Tr}[U_{\vec{x}} \rho \otimes \ket{0} \bra{0} U_{\vec{x}}^{\dagger} \ket{i} \bra{i}]$, and calculate the effects accordingly. More general description of the practical implementation of read-out noise reduction strategies is presented in Appendix \ref{sec:summary_practical_implementation}.

\begin{figure*}[t]
    \includegraphics[width=0.8\textwidth]{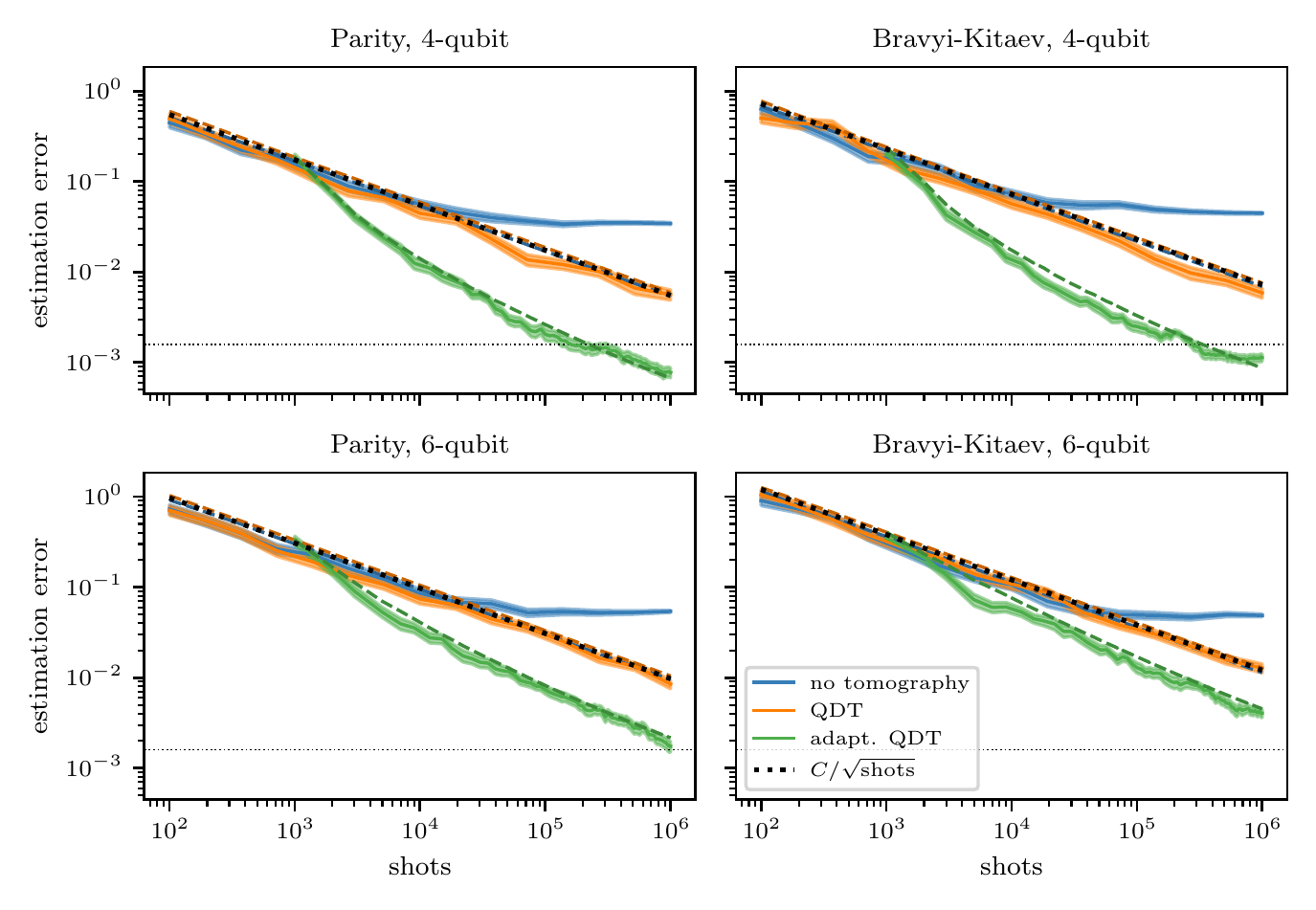}
    \caption{VQE ground state energy estimation error for a simulation of $\rm{H_2}$ molecule with bond length $0.75\textup{~\AA}$, when measurements are performed without a dilation and limiting to PM-simulable POVMs. The computations where performed on a simulated fully connected device, for which the noise parameters were randomly picked from IBM Hanoi system. We present results for parity (left) and Bravyi–Kitaev (right) fermion-to-qubit mappings \cite{mappings} for four (top) and six (bottom) qubits in STO-3G and 6-31G basis, respectively. The blue solid curve gives the true mean error among 40 estimations when no readout noise mitigation is employed. The orange and green solid curves give the true mean errors when QDT is used to correct the bias in the non-adaptive and adaptive strategies respectively. The corresponding areas represent the standard error w.r.t.~true energy of the ground state among 40 runs taken. The dashed lines give the estimated standard errors, that is, computed without knowledge of the true ground energy. The horizontal dotted line represents the chemical accuracy $1.59\textup{~mHa}$, while the oblique dotted line is a reference line $\frac{C}{\sqrt{S}}$ representing a typical estimation error precision decrease with respect to the given number of shots $S$.}
    \label{fig:fake_hanoi_no_dilation}
\end{figure*}
\subsection{Numerical simulations}
\label{sec:numerical_simulations}
Let us now demonstrate the methods presented above on simulated IBM hardware. More specifically, we will compare the convergence of the molecular ground state energy of $\rm{H_2}$ towards chemical precision when characterising the readout noise via QDT. The circuit producing the ground state was obtained with a noiseless Qubit-ADAPT-VQE \cite{q_adapt}. To show the potential of the adaptive measurement strategy of \cite{l2m} in significantly reducing the number of measurement shots in quantum hardware with current noise levels, we compare results for both adaptive and non-adaptive measurement schemes. To allow faster computation times, for the simulations we assumed an IBM Hanoi system extended to full connectivity, such that for missing connections between physical qubits on the real hardware, we randomly drew the noise description of CNOT operations from the existing ones.

When correcting the measurement bias, we started our numerical experiment with detector tomography for each qubit's ideally projective measurement, and in case of dilation strategy, with a CNOT tomography for each gate used in the POVM implementation. For each qubit and gate we used 250,000 shots. Finally, we applied QDT to determine the effects describing the measurement on the system. For each strategy (no mitigation, non-adaptive error mitigation, and adaptive with error mitigation) we repeated the energy estimation process 40 times, with the QDT results introduced before. In the case of adaptive POVMs, the determined QDT results were used as initial step, and every time the POVM parameters were updated we repeated QDT, using 250,000 shots. The 40 results per strategy were used to derive the mean value of the energy and corresponding estimated standard error, presented in Figs.~\ref{fig:fake_hanoi_dilation} and~\ref{fig:fake_hanoi_no_dilation}.

In Fig.~\ref{fig:fake_hanoi_dilation} we show the simulation results with and without measurement bias correction using QDT and compare the non-adaptive and adaptive strategies for the case when the generalised measurements are implemented using a dilation. Each tomography was performed with 250,000 shots. While the simulation without measurement noise mitigation fails to converge to the exact value within chemical precision, we see that both the adaptive and non-adaptive strategies approach this value when the bias is corrected using QDT. We further see that the adaptive measurement protocol offers a clear advantage in terms of measurement shots compared to the standard approach. We would like to note that in the QDT any single qubit gate noise is not taken into account, and whenever such errors become significant the tomography may not be accurate enough to remove the measurement noise bias. However, with the realistic noise model under study here, the single qubit errors seem to be sufficiently small to allow for our assumption to be valid.

For Fig.~\ref{fig:fake_hanoi_no_dilation} we limit the measurements to class of PM-simulable POVMs to see how dropping the use of auxiliary qubits for measurements influences the amount of required measurement shots. We see that restricting the POVMs to the PM-simulable class slows down the convergence especially for the non-adaptive measurement scheme, but hardly influences the amount of necessary measurement shots in the adaptive scheme. Thus we see that even with a dilation-free POVM class the adaptive method clearly outperforms the non-adaptive strategy.

\section{Conclusions and Outlook}
\label{sec:conclusions}

In this paper, we propose two methods allowing adaptive measurement strategies that help overcome the VQE measurement problem, to be implemented in near-term noisy quantum devices with limited qubit number.
To reduce the qubit overhead resulting from implementing generalised measurements required for an adaptive measurement strategy, we suggest to constraint the measurement optimisation to a restricted class of POVMs, which may be simulated with simple projective measurements and classical post-processing. We find that. even if restricting the measurements to a limited class increases the amount of measurement shots, the adaptive strategy still significantly outperforms the standard approach. 
We further demonstrate how readout noise may be mitigated using quantum detector tomography. The numerical simulations performed under realistic noise conditions clearly show that the suggested method removes a bias that would otherwise result from the incorrect assumption on the performed measurements.

Overall, we demonstrate that together the methods presented here significantly reduce the number of needed shots to achieve chemical precision in quantum simulation under noise levels typical to current hardware without qubit overhead due to generalised measurements. The results indicate that adaptive measurement strategies combined with optimised measurement implementations and appropriate noise mitigation can help remove one of the most critical bottlenecks of near-term quantum computing. Our findings highlight how IC POVMs are a key ingredient in unlocking the realistic use of variational algorithms on near-term quantum computers for quantum chemistry simulations, thus paving the way to useful quantum advantage.

The methods presented here may be further optimised for current hardware to reduce the measurement shots beyond what was demonstrated here. The design of a generalised measurement strategy combining both dilation and PM-simulable measurements optimised for specific hardware with limited connectivity remains a topic for future research. Further, as discussed in the previous section, we did not address the influence of single-qubit noise that may deteriorate the accuracy of the QDT when non-negligible, but we leave this as a subject of further study.

\textbf{Competing interests:} Elements of this work are included in patents filed by Algorithm Ltd with the European Patent Office.
\textbf{Authors contributions:} GGP, SM and MR conceived the idea of the methods.
EMB, GGP and ZZ designed and directed the research.
AG, AN, GGP and MR implemented the algorithms and ran the simulations.
EMB, GGP, MR and ZZ wrote the first version of the manuscript.
All authors contributed to scientific discussions and to the writing of the manuscript.

\bibliography{references.bib}

\newpage
\appendix

\section{Practical implementations of dilation-free POVMs on typical cloud-based quantum devices}
\label{sec:reduce_latency}
In this section we discuss a technique to speed up the implementation  of dilation-free POVMs in some specific platforms where the latency in the optimisation loop may a problem.

Suppose that we want to collect IC data for an $N$-qubit state $\rho$ by using the dilation-free POVM method explained in Section \ref{sec:method_description} on each individual qubit composing the state, on a gate-based quantum computer, such as the IBM Quantum devices, accessible from a cloud system.
The state $\rho$ may, e.g., be an ansatz state for a VQE algorithm, for which one wants to estimate the average of an operator such as a Hamiltonian.

Typical cloud-based quantum devices accept a batch of $N_c$ circuits, where a circuit is a specification of a sequence of gates to be applied to one or more of the qubits and ends with the projective measurement on the computational basis $\{\ket{0}, \ket{1}\}$ of one or more qubits.
Each circuit is implemented on the device repeatedly for a given number of shots $S$.
For each shot, the outcomes of the measurements on the different qubits are recorded.

We assume that different POVMs are implemented on each qubit.
Thus, to the $j$-th qubit, we associate a tuple $(\{\alpha^{(j)}_k\}_{k=0}^{K-1}, \{ U^{(j)}_k\}_{k=0}^{K-1})$ and the probability distributions $\lbrace \lbrace P^{(j)}(m | k, b)\rbrace_{m=0}^{M-1} \rbrace_{k=0,b=0}^{K-1,1}$ characterising the POVM. We also assume that the set of operations needed for preparing the $N$ qubits in the state $\rho$ are known. Our method comprises the following steps:

\begin{enumerate}
    \item On a classical computer we construct a batch of circuits to be executed on the quantum device in the following way: we sample a total of $N_S$ length-$N$ lists of integer indices $(k^{(1)}, \ldots, k^{(N)})$, where each $k^{(j)}$ is drawn according to the probabilities $\{ \alpha_k^{(j)}\}_{k=0}^{K-1}$. 
    
    \item For each list, a circuit is defined where, after the initial state preparation, each qubit is rotated according to the chosen unitary $U_k^{(j)}$ before performing a projective measurement in the computational basis.
    
    \item The batch of circuits is executed on the quantum device. Each circuit is executed for a number of shots corresponding to the number of times the corresponding list of indices was sampled. A list of bit strings representing the measurement outcomes is downloaded to the classical computer.
    
    \item For each circuit and for each qubit in the circuit the outcome $b$ is relabeled with a value $m$ sampled from the distribution $P(m|k,b)$. The collection of relabelled strings of outcomes can be used to estimate mean values of observables or other quantities that require IC data.

\end{enumerate}

\vspace{2mm}
\section{Explicit calculation of model effects for read-out noise mitigation for adaptive POVMs}
\label{sec:ModelEffects}
The set of effects $\{ \Pi_{\mathrm{model}, i} (\vec{x}) \}$ fulfilling Eq.\ \eqref{eq:effect_model} can be obtained explicitly.
We assume that we have an explicit expression for the channel $\mathcal{E}_{\vec{x}}$, so that given some operator $\mathcal{O}$, we can calculate $\mathcal{E}_{\vec{x}} (\mathcal{O})$ classically. We also assume we have obtained the effects $\{ M_{\mathrm{QDT}, i} \}$ through QDT.

It is convenient to introduce an orthonormal basis $\{ B_a \}$ for the space of linear operators $L(\mathcal{H}_{\mathrm{s}})$, where $\mathcal{H}_{\mathrm{s}}$ is the Hilbert space of the subsystem measured via the POVM.
These operators satisfy the relation $\mathrm{Tr} [B_a B_b] = \delta_{ab}$.

Expressing the density matrix $\rho = \sum_a \rho_a B_a$ and the effect $\Pi_{\mathrm{model}, i} (\vec{x}) = \sum_a \pi_a B_a$ in this basis, the left-hand side of Eq.\ \eqref{eq:effect_model} reads $\mathrm{Tr}[ \rho \Pi_{\mathrm{model}, i} (\vec{x})] = \sum_a \rho_a \pi_a$.
Introducing the decomposition in the right-hand side instead yields $\mathrm{Tr}[\mathcal{E}_{\vec{x}} (\sum_a \rho_a B_a \otimes \rho_{\mathrm{QST}}) M_{\mathrm{QDT}, i}] = \sum_a \rho_a \mathrm{Tr}[\mathcal{E}_{\vec{x}} (B_a \otimes \rho_{\mathrm{QST}}) M_{\mathrm{QDT}, i}]$, which implies that $\pi_a = \mathrm{Tr}[\mathcal{E}_{\vec{x}} (B_a \otimes \rho_{\mathrm{QST}}) M_{\mathrm{QDT}, i}]$ for the equality to hold for any $\rho$.
Thus, the solution is given by
\begin{equation}\label{eq:effect_model_solution}
    \Pi_{\mathrm{model}, i} (\vec{x}) = \sum\limits_a \mathrm{Tr}[\mathcal{E}_{\vec{x}} (B_a \otimes \rho_{\mathrm{QST}}) M_{\mathrm{QDT}, i}] B_a.
\end{equation}

We can verify that this results is a proper POVM.
First, the operators $\Pi_{\mathrm{model}, i} (\vec{x})$ are positive semidefinite, that is, $\bra{\psi} \Pi_{\mathrm{model}, i} (\vec{x}) \ket{\psi} \geq 0, \, \forall \ket{\psi} \in \mathcal{H}_{\mathrm{s}}$:
\begin{widetext}
\begin{equation}
    \begin{aligned}
        \bra{\psi} \Pi_{\mathrm{model}, i} (\vec{x}) \ket{\psi} &= \mathrm{Tr} \left[\sum\limits_a \mathrm{Tr}[\mathcal{E}_{\vec{x}} (B_a \otimes \rho_{\mathrm{QST}}) M_{\mathrm{QDT}, i}] B_a \ket{\psi} \bra{\psi} \right] = \sum\limits_a  \mathrm{Tr}[\mathcal{E}_{\vec{x}} (B_a \otimes \rho_{\mathrm{QST}}) M_{\mathrm{QDT}, i}] \mathrm{Tr} [ B_a \ket{\psi} \bra{\psi}] \\
        &= \mathrm{Tr} \left[\mathcal{E}_{\vec{x}} \left(\sum\limits_a \mathrm{Tr} [ B_a \ket{\psi} \bra{\psi} ] B_a \otimes \rho_{\mathrm{QST}} \right) M_{\mathrm{QDT}, i} \right] = \mathrm{Tr} \left[\mathcal{E}_{\vec{x}} (\ket{\psi} \bra{\psi} \otimes \rho_{\mathrm{QST}} ) M_{\mathrm{QDT}, i} \right] \geq 0.
    \end{aligned}
\end{equation}
\end{widetext}
The last expression is non-negative because $M_{\mathrm{QDT}, i}$ are positive semidefinite by definition (they are POVM effects describing the noisy native measurement), and $\mathcal{E}_{\vec{x}} (\ket{\psi} \bra{\psi} \otimes \rho_{\mathrm{QST}})$ is positive given the complete positivity of the channel.
Notice that in the previous to last step we have used the fact that $\ket{\psi} \bra{\psi} = \sum_a \mathrm{Tr} [ B_a \ket{\psi} \bra{\psi} ] B_a$, which follows from the orthogonality of the basis operators: if $\ket{\psi} \bra{\psi} = \sum_b \psi_b B_b$, then $\mathrm{Tr} [ B_a \ket{\psi} \bra{\psi} ] = \sum_b \psi_b \mathrm{Tr} [ B_a B_b ] = \psi_a$.

We can also verify that the effects add up to identity, i.e.\ $\sum_i \Pi_{\mathrm{model}, i} (\vec{x}) = \mathbb{I}$.
To simplify the calculations, let us fix one of the basis elements, for instance the zeroth, $B_0 = \mathbb{I} / \sqrt{d}$, 
where $d = \dim (\mathcal{H}_{\mathrm{s}})$.
By doing so, the orthogonality condition, $\mathrm{Tr} [B_0 B_a ] = \delta_{0a}$, implies that all other basis elements must be traceless, $\mathrm{Tr} [ B_a ] = \sqrt{d} \delta_{0a}$.
The direct calculation of the sum of the operators in Eq.\ \eqref{eq:effect_model_solution} yields
\begin{equation}
    \begin{aligned}
    \sum\limits_i \Pi_{\mathrm{model}, i} (\vec{x}) &= \sum\limits_i \sum\limits_a \mathrm{Tr}[\mathcal{E}_{\vec{x}} (B_a \otimes \rho_{\mathrm{QST}}) M_{\mathrm{QDT}, i}] B_a \\
    &= \sum\limits_a \mathrm{Tr} \left[\mathcal{E}_{\vec{x}} (B_a \otimes \rho_{\mathrm{QST}}) \sum\limits_i M_{\mathrm{QDT}, i} \right]  B_a \\
    &= \sum\limits_a \mathrm{Tr}[\mathcal{E}_{\vec{x}} (B_a \otimes \rho_{\mathrm{QST}})] B_a = \sum\limits_a \mathrm{Tr}[B_a] B_a \\
    &= \sum\limits_a \sqrt{d} \delta_{0a} B_a = \mathbb{I}.
    \end{aligned}
\end{equation}
We have used the fact that $\sum_i M_{\mathrm{QDT}, i} = \mathbb{I}$, since it is a POVM, that $\rho_{\mathrm{QST}}$ is trace-one, and that $\mathcal{E}_{\vec{x}}$ is trace-preserving.

\section{Practical implementation of measurement error mitigation strategies}\label{sec:summary_practical_implementation}
The ideas presented here for measurement error mitigation can be summarised by grouping them into two different methods, the practical implementation steps of which are listed below.

\subsection{Measurement error mitigation through QDT}\label{sec:method_1}

Given a sequence of physical operations targeted at implementing a POVM, which can be a specific instance of a parametric family of POVMs or otherwise a parameter-free POVM, we characterise mathematically the quantum mechanical measurement resulting from such implementation in the physical device through QDT.
The obtained description thus takes into account, at least partly, the impact of noise and imperfections on the probability distribution of outcomes, and it is then used to estimate expectation values of observables.
The resulting estimations are expected to have smaller biases due to measurement errors than those obtained using other measurement error mitigation methods.

\subsubsection{Implementation steps}

The steps to implement the method on a multi-qubit system on which single-qubit POVMs are used are the following.

\begin{itemize}
    \item[1.] Perform QDT of the implemented POVM on every qubit in parallel.
    This generally requires preparing all the qubits in an informationally complete set of states and applying the POVM measurement on them repeatedly to gather outcome statistics; these are then used to reconstruct the POVM effects $\{ \Pi_{\mathrm{QDT}, m}^{(i)} \}$ of each qubit $i$ on a classical computer.
    \item[2.] For every qubit $i$, determine if the reconstructed effects $\{ \Pi_{\mathrm{QDT}, m}^{(i)} \}$ form a basis of $L(\mathcal{H}_{\mathrm{s}})$. In practice, this implies finding a subset of $d^2$ linearly independent effects.
    If such subset exists, proceed to the next steps.
    Otherwise, the implemented measurement is not informationally complete.
    \item[3.] For every qubit $i$, find the decomposition of the Pauli operators $\sigma_{k}^{(i)}$ in terms $\sigma_{k}^{(i)} = \sum_m b_{km}^{(i)} \Pi_{\mathrm{QDT}, m}^{(i)}$.
    \item[4.] Use the POVM to measure on the quantum state $\rho$ of interest, that is, prepare it on the device repeatedly and measure each qubit using the required sequence of POVM operations.
    Use the hybrid quantum-classical Monte Carlo method from Ref.\ \cite{l2m}, computing the $\omega_{\mathbf{m}}$ with the $b$-matrix elements from step 3, to estimate the expectation value of observables of interest.
\end{itemize}

\subsection{POVM optimisation under experimental imperfections}\label{sec:method_2}

If the sequence of physical operations involves some that are parameterisable, the different sequences corresponding to different parameter values can generally lead to different POVMs in the real device.
Seizing the informational completeness of the outcomes to optimise the POVM without measurement overhead, in a similar manner to Ref.\ \cite{l2m}, requires a model of the implemented POVM, such that we can classically predict the POVM effects resulting from the physical implementation of the POVM for any given parameter values.
While the model is useful for finding better POVMs classically, using the method from Sect.~\ref{sec:method_1} to characterise the POVM once a set of parameters is proposed may yield more accurate results.
In other words, the measurement method in Sect.~\ref{sec:method_1} should be preferably used for observable averaging; the method in this section is mainly aimed at exploring the POVM space efficiently.
In any case, if the detector model is accurate, the use of the method in Sect.~\ref{sec:method_1} at every optimisation iteration may be unnecessary.

\subsubsection{Implementation steps}
The steps to optimise the real POVM in an imperfect device are the following.

\begin{enumerate}
    \item For each qubit $i$, construct a model of the implemented POVM family.
    Such model consists of a classical computation that yields a set of effects, $\{ \Pi_{\mathrm{model}, m}^{(i)} (\vec{x}_i) \}$, where $\vec{x}_i$ is the set of parameters associated with qubit $i$'s POVM, that provide an accurate description of the real measurement process.
    Constructing such model requires the following.
    \begin{enumerate}
        \item A model of the physical measurement taking place at the end of the process on all the quantum degrees of freedom involved in the POVM implementation of the qubit.
        This model is composed of a collection of POVM effects $\{ M_{\mathrm{QDT}, m}^{(i)} \}$ reconstructed through QDT in a separate set of experiments. Notice that these effects are different from $\{ \Pi_{\mathrm{QDT}, m}^{(i)} (\vec{x}_i) \}$, as they only describe the native measurement of the device, not including any other operations in the POVM implementation.
    
        \item A model of the physical transformation, i.e.\ the quantum channel, resulting from the sequence of operations before the final measurement takes place.
        The classical description of the channel $\mathcal{E}_{\vec{x}_i}^{(i)}$ may be obtained as a composition of sequential channels, some of them parameter-free, and others with parameter dependence.
        Their classical description may be given in terms of e.g.\ their Choi matrix, or through a master equation numerically integrated.
        
        \item A model of the state of the ancillas right before the POVM unitary is applied, $\rho_{QST}$, obtained by means of single-qubit QST.
    \end{enumerate}
    
    With these elements, the effects that approximate the description of the real measurement apparatus for any $\vec{x} = (\vec{x}_1, \ldots, \vec{x}_n)$ can be obtained using Eq.\ \eqref{eq:effect_model_solution}.
    \item Choose an initial value $\vec{x}$ for the parameters.
    \item Implement the method in Sect.~\ref{sec:method_1}, including applying the measurement to the state of interest $\rho$ and the estimation of observables, for the POVM corresponding to parameters $\vec{x}$.
    The reason why this step may be useful is that we assume that $\{ \Pi_{\mathrm{model}, m}^{(i)} (\vec{x}_i) \}$ gives a less accurate description of the physical measurement process than $\{ \Pi_{\mathrm{QDT}, m}^{(i)} (\vec{x}_i) \}$.
    Thus, we may rely on $\{ \Pi_{\mathrm{model}, m}^{(i)} (\vec{x}_i) \}$ to assess the performance of POVMs classically, as in step 4 below, but $\{ \Pi_{\mathrm{QDT}, m}^{(i)} (\vec{x}_i) \}$ are expected to yield more accurate estimations of the observable.
    Yet, we have in principle no way to determine $\{ \Pi_{\mathrm{QDT}, m}^{(i)} (\vec{x}'_i) \}$ classically for $\vec{x}'_i$ that have never been used, while the model effects $\{ \Pi_{\mathrm{model}, m}^{(i)} (\vec{x}'_i) \}$ serve precisely that purpose.
    \item Using the IC data from the previous measurement rounds, find parameter values $\vec{x}'_i$ for which the corresponding $\{ \Pi_{\mathrm{model}, m}^{(i)} (\vec{x}'_i) \}$ are expected to perform better at estimating the observable of interest.
   This can be done following a similar procedure to the one in Ref.\ \cite{l2m}, Sect. III B.
    Essentially, we need to be able to estimate classically the second moment of the Monte Carlo sampling for POVMs with parameters $\vec{x}'$ for which the POVM may have not been used on the device.
    
    Suppose that we have measured the state (step 3) with POVM parameters $\vec{x}$, for which the method in step 3 previously inferred $\{ \Pi_{\mathrm{QDT}, m}^{(i)} (\vec{x}_i) \}_{i, m}$.
    To optimise the POVM, we need to estimate $\langle \omega_{\mathbf{m}} (\vec{x}')^2 \rangle$ classically.
    As explained in Ref.\ \cite{l2m}, this can be done efficiently if only one qubit's POVM is different for $\vec{x}'$.
    In general, however, it is possible to calculate this classically if more qubits' POVMs are modified, but the classical overhead of the calculation grows exponentially in the number of modified single-qubit POVMs.
    
    For simplicity, let us assume that $\vec{x}'_i = \vec{x}_i $ for all $i \neq q$.
    
    \begin{enumerate}
        \item  We decompose qubit $q$'s new effects, $\{ \Pi_{\mathrm{model}, m}^{(q)} (\vec{x}'_q) \}$, in terms of the (reliable) tomographically reconstructed old ones, $\Pi_{\mathrm{model}, r}^{(q)} (\vec{x}'_q) = \sum_m d_{rm}^{(q)} \Pi_{\mathrm{QDT}, m}^{(q)} (\vec{x}_q)$.
        \item  Equipped with this $d$-matrix, we can reuse the IC data from the previous step to estimate $\langle \omega_{\mathbf{m}} (\vec{x}')^2 \rangle$ as
    \begin{equation}
        \bar{\omega}_{\mathbf{m}}^2 (\vec{x}') = \frac{1}{S} \sum\limits_{\mathbf{m} \in \{ \mathbf{m}_1, \ldots, \mathbf{m}_S \}} \sum\limits_{r_q = 0}^{M_q} d_{r_q m_q}^{(q)} \left[\omega_{(m_1, \ldots, r_q, \ldots, m_n)} (\vec{x}') \right]^2.
    \end{equation}
    \end{enumerate}
    
    Notice that $\omega_{(m_1, \ldots, r_q, \ldots, m_n)}$ relies on the $b$-matrix for qubit $q$ corresponding to $\{ \Pi_{\mathrm{model}, m}^{(q)} (\vec{x}'_q) \}$.
    Also, the $d_{r_q m_q}^{(q)}$ implicitly depend on $\vec{x}'_q$ through the decomposition of the model effects.
    The above quantity can be minimised for $\vec{x}'_q$ on a classical computer to obtain the parameters $\vec{x}^{\textrm{next}}_q$ for the next iteration, for which the POVM should be more efficient.
    This can be done using gradient descent or any other method.
    The same operation can be carried out for every qubit $i$, to produce $\vec{x}^{\textrm{next}} = (\vec{x}^{\textrm{next}}_1, \ldots, \vec{x}^{\textrm{next}}_n)$.
    \item Set $\vec{x} = \vec{x}^{\textrm{next}}$ and go to step 3 unless convergence of the adaptive POVM algorithm has been reached.
\end{enumerate}

As in Ref.\ \cite{l2m}, at the end of the process, the IC data has been used to evaluate one estimator $\bar{\mathcal{O}}_t$ per iteration $t$, as well as its variance $\bar{V}_t$. These estimators can all be put together using the on-the-fly method in that reference.

The detector tomography in step 3 above can be skipped and the model effects $\{ \Pi_{\mathrm{model}, i} (\vec{x}) \}$ can be used instead of $\{ \Pi_{\mathrm{QDT}, i} (\vec{x}) \}$ if they can be considered accurate enough.

\end{document}